\documentclass{emulateapj}
\usepackage{graphicx}
\usepackage{bm}
\usepackage{amssymb}
\usepackage{amsmath}
\usepackage{cancel}
\usepackage{hyperref}
\usepackage{physics}

\usepackage{savesym}
\savesymbol{tablenum}

\usepackage[dvipsnames]{xcolor}
\usepackage{multirow}

\usepackage{epstopdf}
\usepackage{natbib}
\usepackage{epsfig}
\usepackage{cleveref}
\usepackage{color}

\newcommand{\beq}{\begin{equation}}
\newcommand{\eeq}{\end{equation}}

\newcommand{\tot}{\text{tot}}
\newcommand{\kms}{\text{km}/\text{s}}
\newcommand{\pc}{\text{pc}}
\newcommand{\kpc}{\text{kpc}}
\newcommand{\Myr}{\text{Myr}}
\newcommand{\Gyr}{\text{Gyr}}
\newcommand{\actunit}{\kpc\,\kms}
\newcommand{\feh}{[\text{Fe}/\text{H}]}
\newcommand{\mgfe}{[\text{Mg}/\text{Fe}]}
\newcommand{\alphafe}{[\alpha/\text{Fe}]}
\newcommand{\comma}{\text{,}\,}
\newcommand{\ms}{\text{m}/\text{s}}

\newcommand{\ts}{\textsuperscript}

\usepackage{hyperref}

\begin{document}
\title{Actions are weak stellar age indicators in the Milky Way disk}
\author{Angus Beane\altaffilmark{1,2}, Melissa K. Ness\altaffilmark{2,3}, \& Megan Bedell\altaffilmark{2}}
\altaffiltext{1} {Department of Physics \& Astronomy, University of Pennsylvania, 209 South 33rd Street, Philadelphia, PA 19104, USA}
\altaffiltext{2} {Center for Computational Astrophysics, Flatiron Institute, 162 5th Ave., New York, NY 10010, USA}
\altaffiltext{3} {Department of Astronomy, Columbia University, 550 W 120th St, New York, NY 10027, USA}
\email{abeane@sas.upenn.edu}
\begin{abstract}
The orbital properties of stars in the disk are signatures of their formation, but they are also expected to change over time due to the dynamical evolution of the Galaxy. Stellar orbits can be quantified by the three dynamical actions, $J_r$, $L_z$, and $J_z$, which provide measures of the orbital eccentricity, guiding radius, and non-planarity, respectively. 
Changes in these dynamical actions over time reflect the strength and efficiency of the evolutionary processes that drive stellar redistributions. We examine how dynamical actions of stars are correlated with their age using two samples of stars with well-determined ages: 78 solar twin stars (with ages precise to $\sim5\%$) and 4376 stars from the APOKASC2 sample ($\sim20\%$). We compute actions using spectroscopic radial velocities from previous surveys and parallax and proper motion measurements from {\em Gaia DR2}. We find weak gradients with significant scatter for all actions as a function of stellar age. These gradients and their associated variances provide strong constraints on the efficiency of the mechanisms that drive the redistribution of stellar orbits over time and demonstrate that actions are informative as to stellar age. However, the shallow action-age gradients combined with the large dispersion in each action at a given age renders the prospect of age inference from orbits of individual stars bleak. Using the precision measurements of $\feh$ and [$\alpha$/Fe] we find that similarly to our stellar age results, the dynamical actions afford little discriminating power between individual low- and high-$\alpha$ stars.

\end{abstract}

\section{Introduction}

Secular evolution is the process by which stars redistribute on time scales longer than a galaxy's dynamic time scale (for a recent review, see \citet{Sellwood14:secreview}). An important process in secular evolution is radial mixing, in which stars are redistributed through changes in angular momentum, which alters the radius of their orbit. Radial mixing tends to increase the eccentricity of a star's orbit, but not always. This redistribution can be caused by several processes, and is important for understanding the chemical landscape and other features of the Milky Way disk \citep{Rovskar08,Schonrich09}.

The presence of a non-axisymmetric disturbance (e.g. spiral arms and bars) is the dominant factor behind radial mixing. 
Non-axisymmetric perturbations provide a driving force which, when in resonance with a star's orbit, can increase or decrease the angular momentum of the star's orbit.
It has long been understood that radial mixing occurs near the inner Lindblad resonance, causing stars to move outwards, and vice versa for an outer Lindblad resonance \citep{LyndenBell72}. In both cases the eccentricity of a star's orbit is increased. It was later shown that stars orbiting at corotation resonance also exhibit significant radial mixing, but without significant changes to eccentricity or vertical heating, referred to as radial migration \citep{Sellwood02:radialmixing}.\footnote{Here, by radial migration we only mean angular momentum redistribution at corotation resonances --- not the expected redistribution of angular momentum from any non-axisymmetric disturbance, which we refer to as ``radial mixing,'' following \citet{Sellwood14:secreview}.} \citet{LyndenBell72} also showed that angular momentum transfer occurs at corotation resonances, though the importance of corotation resonances was not appreciated until later.

Other, more complicated processes can also drive radial mixing. In the presence of multiple non-axisymmetric perturbations (e.g. bars and spirals), resonance overlap can induce even larger changes in angular momentum \citep{Minchev10:resonanceoverlap,Minchev11:resonanceoverlapsim}. Satellite bombardment has also been shown to induce radial mixing \citep{Bird12:satellites}.

While interactions between spiral arms and stars can drive evolution in orbital radius, the mechanism behind vertical heating is less certain. Spiral arms are not expected to drive vertical heating as the typical vertical oscillation frequency $\Omega_z$ is much larger than the spiral arm perturbation frequency $\Omega_p$ to be in resonance \citep{Carlberg87:vheating,Loebman11:vheating,Minchev12:vheating}.
Studying the extent to which spiral arms thicken in simulation is difficult, as the thickening is influenced by the specifics of gravity softening \citep{Sellwood13:vheatingsim,Solway12:vheatingsim}. Recent work has shown that in thin disk only models giant molecular clouds (GMCs) are necessary in order to reproduce the age-velocity dispersion relation seen in the solar neighborhood \citep{Aumer16:paper1,Aumer16:paper2}. However, when the thick disk is included, less GMC heating is needed to reproduce observations \citep{Aumer17:paper4}.

 Since radial mixing processes occur over fairly long time-scales, we expect the dynamical properties of stars to be, in a sense, dynamical clocks. It has long been understood that older stars have larger velocity dispersion than younger stars \citep{Wielen77:dispersion_age,Nordstrom04:GenevaCopenhagen}. In the pre-{\em Gaia} era, the best evidence for this relationship came from the radial velocity estimates of the Geneva-Copenhagen survey \citep{Nordstrom04:GenevaCopenhagen}, with follow-up measurements of stellar parameters and ages \citep{Holmberg07:GCAges} and revised parallax and proper motion estimates from {\em Hipparcos} \citep{vanLeeuwen07:revisedHipparcos}. Using this data, it has been shown that the dispersion for the $U$, $V$, and $W$ velocities increases with age, with the most probable conclusion being that stellar scattering increases the velocity dispersion over time \citep{Holmberg09:disp_age,Casagrande11:disp_age}. These studies have been questioned with respect to the accuracy of the stellar age determinations \citep{Soderblom10:ages_are_bad,Reid07:ages_are_bad}.

This paper is an improvement on previous studies of stellar scattering for several reasons. First, we use two recent samples of stars (a solar twin sample and the APOKASC2 sample) with much better determined ages than has been previously available. Second, by using the improved astrometry of {\em Gaia DR2}, we of course are able to measure the dynamical properties of stars more accurately. But, most importantly, we present results in action space instead of velocity dispersion space. This offers a completely new perspective, which we discuss below.

\begin{figure}[t!]
\centering
\includegraphics[width=\columnwidth]{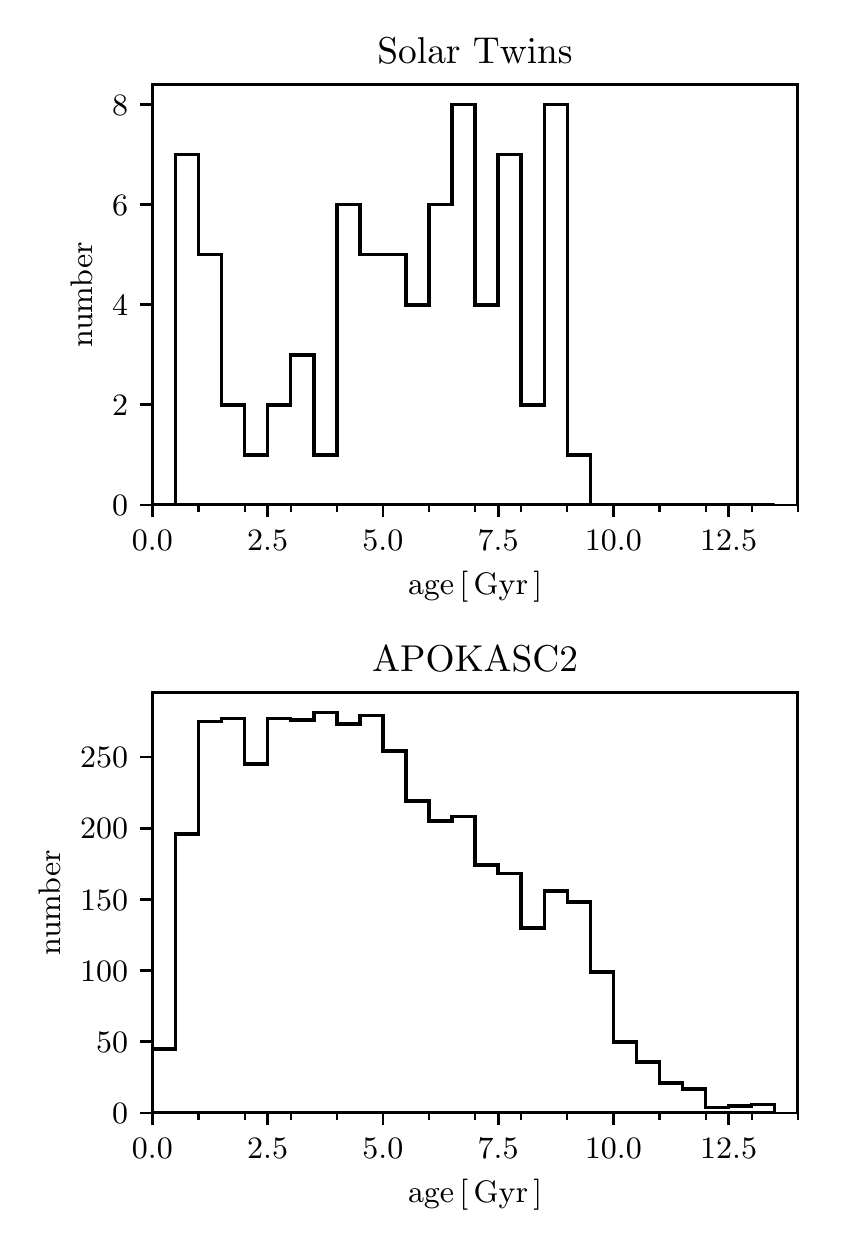}
\label{fig:age_hist}
\caption{Age histograms of the two samples of stars we use in this study. The bins are linearly spaced with widths of $0.5\,\Gyr$. The APOKASC2 sample is biased towards younger stars, while the solar twin sample contains a more uniform distribution, with $\sim6\,\Gyr$ stars slightly favored.}
\end{figure}

Action space provides an excellent way to characterize stellar orbits. The three actions have intuitive physical meanings, and in the case of a slowly-evolving axisymmetric potential they are both conserved and completely characterize a star's orbit.  Thus, any evolution in action space is caused by the non-axisymmetric nature of the Milky Way. The three actions are computed in the phase space of the 3D positions of stars in cylindrical coordinates: $(r\text{,}\,\phi\text{,}\,z)$, and the relevant conjugate momenta. They are defined as,
\beq\label{eq:actions}
J_i \equiv \frac{1}{2\pi} \oint_{\text{orbit}}p_i\,\text{d}x_i,
\eeq
where $i=r\text{,}\,\phi\text{,}\,z$. $J_r$ quantifies the radial excursions of an orbit, and vanishes for circular orbits. $J_z$ quantifies the vertical excursions of an orbit, and vanishes for orbits lying entirely in the galactic plane. $J_{\phi}$ is equivalent to a star's orbital angular momentum, $L_z$. We will refer to $L_z$ throughout this work. For a more detailed introduction to action space, see e.g. \citet{Sellwood14:secreview,Binney08}.

In some sense, actions and velocity dispersion measure the same quantity. For example, the vertical action $J_z$ measures the vertical excursions of an orbit. The same information would be included in the vertical velocity dispersion $\sigma_W$ for a population of stars on the same orbit. This is because an orbit with higher vertical excursions will have greater vertical velocity as the star passes through the galactic plane, increasing $\sigma_W$. However, given accurate enough astrometry, an action can be calculated for a single star. Thus, the effects of spiral arm scattering, bar scattering, cloud scattering, etc. can be understood at the level of {\em individual stars}, rather than populations of stars, a necessary restriction of previous work. Furthermore, if actions and ages are strongly correlated then one can in principle invert the relation to infer an individual star's age from its actions.

It has been argued that the Sun underwent a radial migration process, traveling from a smaller galactic radius (estimates vary from $5 \text{--} 7\,\kpc$) to its current $8\,\kpc$.
Arguments have been made based on the metallicity gradient of the Milky Way and the radial dispersion around this gradient as a function of age, as well as the observation that the Sun is more metal-rich than nearby stars of similar age \citep{Wielen96:solarbirth,Holmberg09:chemsurvey,Minchev14:chemsim,Minchev18:chemgradient,Frankel18:sunmigration}. However, the higher relative metallicity of the Sun compared to the solar neighborhood has been called into question \citep{Casagrande11:chemsolnb,Gustafsson98,Gustasson08,Gustafsson10}. In addition, solar orbit integrations have suggested that migration from the inner galaxy is unlikely, even when uncertainties in the galactic potential and solar position and velocity are taken into account \citep{Martinez14}.

The chemical composition of stars provide markers of stellar birth properties. For instance, older stars tend to have higher $\alpha$-enrichment due to the time delay of SN Ia compared to SN II \citep{Tinsley79}. Because of this, $\alpha$-enrichment is expected to be related to the secular evolution of the Milky Way disk. For instance, the Milky Way disk shows a bimodal $\alpha$-sequence.  Many previous studies have shown that the kinematically and chemically defined thin and thick disks in the solar neighborhood, which comprise the low- and high-$\alpha$ sequence respectively, have different properties. This includes, for example the different metallicity gradients of these populations, which is presumably a consequence of different formation and evolutionary histories \citep{Edvardsson93:thickdisk,Prochaska00:thickdisk,Bensby03:alphabimodal,Ishigaki12:alphathick,Duong18:galah}. 

In this work we examine the relationship between orbital actions and both age and chemical enrichment for stars of the Milky Way disk. By doing so, we offer insight into stellar scattering at the level of individual stars. By working in action space and by using recent samples of stars with well-determined ages, our work is an improvement on previous studies of stellar scattering. In \S~\ref{sec:methods}, we describe the two samples of stars with precisely measured ages that we use to perform this analysis and the details of our action calculations. In \S~\ref{sec:results} we present our results and discuss in \S~\ref{sec:discussion} before concluding in \S~\ref{sec:conclusions}. We assume for the solar position that $R_{\sun} = 8\,\kpc$, $z_{\sun} = 0.025 \,\kpc$, and $\phi_{\sun} = 0$, and for the solar velocity $(U_{\sun}\comma V_{\sun}\comma W_{\sun}) = (11.1\comma 232.24\comma 7.25)\,\kms $. We assume a Local Standard of Rest (LSR) velocity of $220\,\kms$. These assumptions are in agreement with recent results \citep{Juric08:solarposition,Schonrich10:lsr}. 

\begin{figure}[t!]
\centering
\includegraphics[width=\columnwidth]{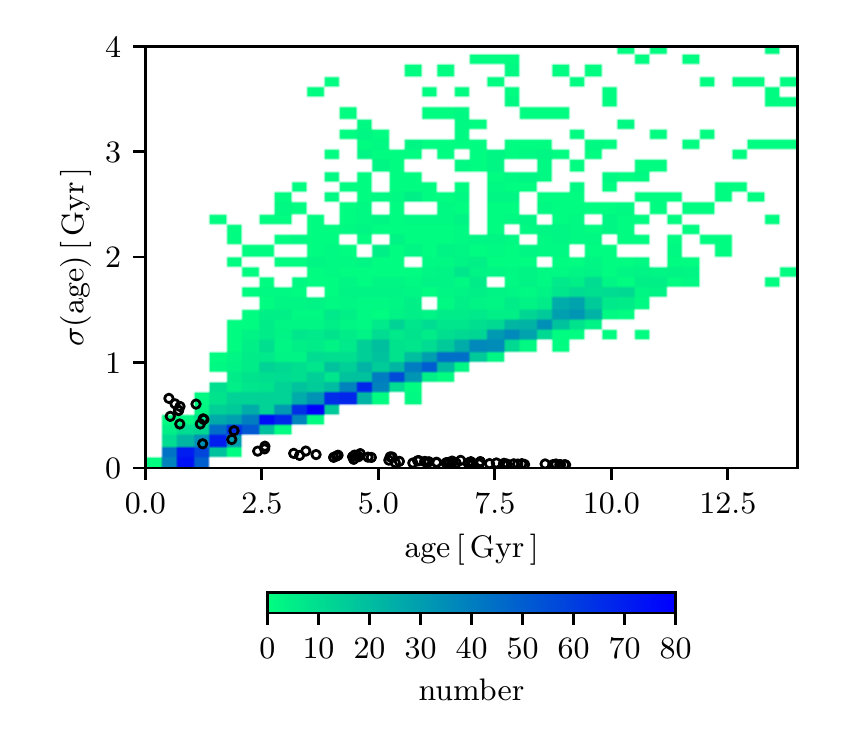}
\caption{The distribution of age uncertainties for our two samples. For the APOKASC2 sample, we took the maximum of the upper and lower uncertainties. A total of $62\%$ of the APOKASC2 sample has age uncertainties less than $20\%$. The solar twin sample has much more precisely determined ages. We have examined the implications of excluding stars with high age errors and verified that using only stars with uncertainties of $<$ $20\%$ in our analysis does not change our results.}
\label{fig:ageunc}
\end{figure}

\begin{figure*}[t!]
\centering
\includegraphics[width=\textwidth]{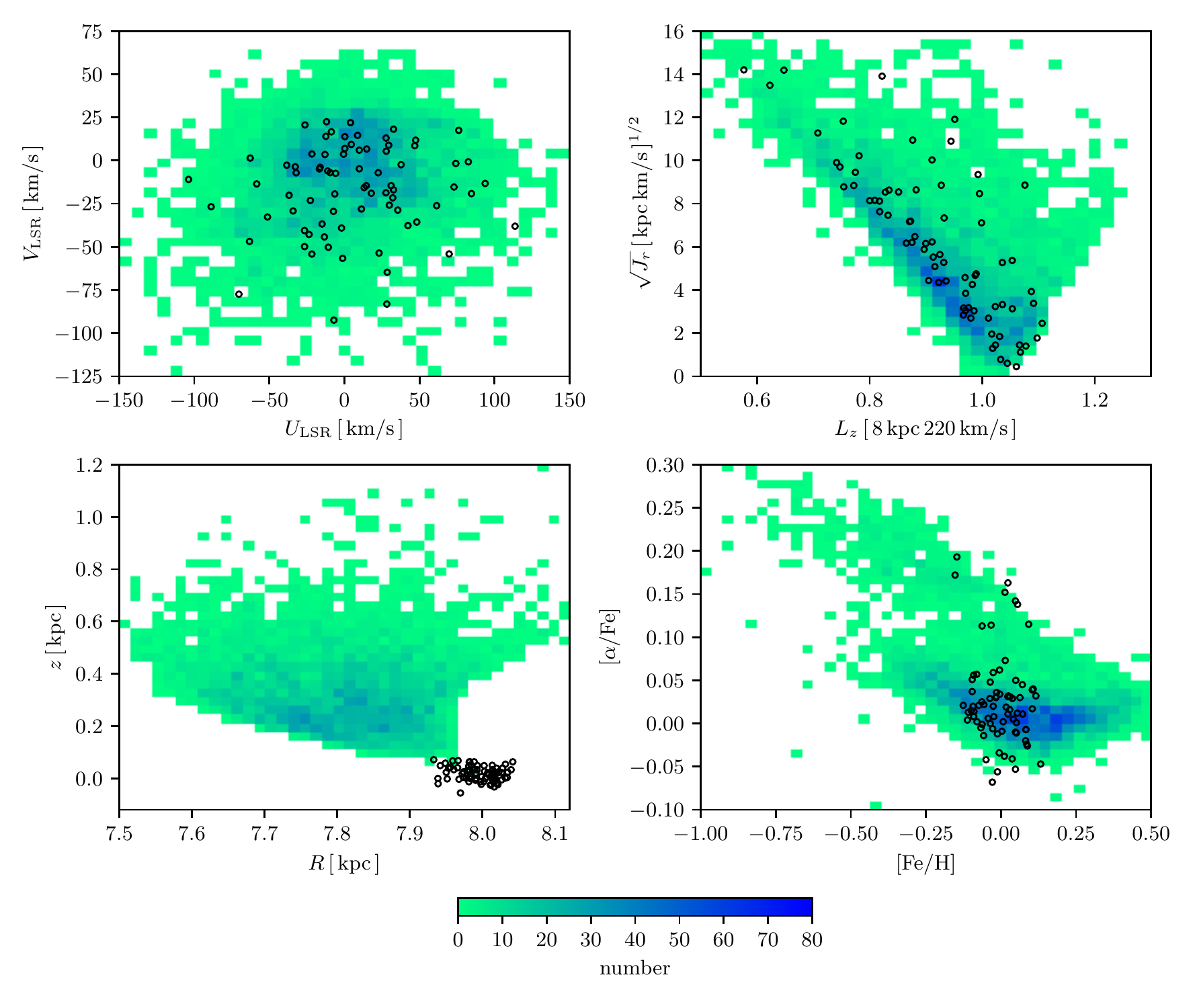}
\caption{The characterization of our two stellar samples in velocity, action, and metallicity space. In all panels, the heatmap corresponds to the APOKASC2 sample while the open circles shows the solar twin sample. The {\em upper left panel} shows the distribution of our stars in the $V$-$U$ plane. One can see the slight trend that stars with higher $V$ tend to also have higher $U$. The {\em upper right panel} shows the distribution of our stars in the $J_r$-$L_z$ plane. The asymmetry and cone-like nature is well-understood and explained in the text. The {\em bottom right panel} shows our distribution of our stars in the $\alphafe$-$\feh$ plane. For the solar twin sample, we are plotting $\mgfe$. Our sample includes both the high and low alpha sequences although the majority of our stars are in the low-alpha sequence ($89\%$ have $\alphafe < 0.12$), which is a consequence of the spatial selection of these stars \citep{Hayden15:chemicalcartography}. The {\em bottom left panel} shows the spatial distribution of each sample. Note that the APOKASC2 sample is biased against stars lying close to the galactic plane.}
\label{fig:char}
\end{figure*}

\section{Data and Methods}\label{sec:methods}
\subsection{Sample Selections \& Quality Cuts}

Our first and most precisely measured sample of stars consists of 78 solar twins, all located within $100\,\pc$ of the Sun. These stars were selected from a variety of large surveys and subsequently followed up with high-resolution spectroscopy to ensure that their fundamental parameters ($T_{\text{eff}}$, $\log{g}$, $[\text{Fe}/\text{H}]$) were close to the solar values at high confidence \citep{ST14}. Stacked spectra from HARPS with a combined signal-to-noise of $\approx 1000$ were collected for each of the 78 solar twins and analyzed using a strict differential approach \citep{Spina18,Bedell18}. The resulting precise stellar parameters provide, in combination with theoretical isochrones, measured ages to $\sim5\%$ uncertainty for each solar twin in the sample and $\alpha$-element abundances (as measured by [Mg/Fe]) with errors $< 0.04\,\text{dex}$.

The recently released APOKASC2 sample also provides precise age measurements for 6676 red giant and red clump stars \citep{Pinsonneault18:apokasc2} . This sample uses a combination of APOGEE spectroscopic parameters and {\em Kepler} asteroseismic data to determine the ages of these stars. The selection for APOKASC2 is complicated, and the reader is referred to \citet{Pinsonneault18:apokasc2} and references therein for more information. Because of the complicated APOKASC2 selection function, we are not able to infer how our results generalize to the underlying stellar population. The APOKASC2 sample has associated $\feh$ and $\alphafe$ measurements precise to $\lesssim 0.04\,\text{dex}$ and $\lesssim 0.02\,\text{dex}$ respectively \citep{GarciaPerez16:ASPCAPchem}.

The age distribution of each sample is shown in Fig.~\ref{fig:age_hist}. There are more young stars than old stars in the APOKASC2 sample. The age uncertainties of both samples is shown in Fig.~\ref{fig:ageunc}. The heatmap corresponds to the age uncertainties of the APOKASC2 sample while the open circles for the solar twin sample. For the APOKASC2 sample, we took the maximum of the upper and lower uncertainty values. One can see that, especially for older stars, the uncertainties in the solar twin sample are quite low. The uncertainties in the APOKASC2 sample are somewhat larger, although still $62\%$ of the sample has age uncertainties below $20\%$. We do not make any cuts on age uncertainties, although we checked that selecting stars with age uncertainties of $<$ $20\%$ does not qualitatively change our results. While noisy age estimates will in principle decrease correlations between actions and ages, the spread we find in the correlations is large enough to be robust to this effect --- though the true underlying relation may still be tighter than what we find.

The radial velocities for the solar twin sample are taken from \citet{ST14}, which used the HARPS spectrograph \citep{Mayor03:harps}. The radial velocities for the APOKASC2 samples are taken from APOGEE \citep{APOGEE17}. Radial velocities are precise to $\lesssim 100\,\ms$ for the APOKASC2 sample \citep{Nidever15:APOGEErv} and $\lesssim 800\, \ms$ for the solar twin sample \citep{ST14}.

Astrometric quantitites (RA, Dec, proper motions, and parallax) are taken from {\em Gaia DR2} \citep{GAIA18, GAIAastrometry18}. Precise distance and velocity measurement are necessary to make precise measurements of actions \citep{Coronado18:disterror}. We adopt the distance and velocity quality cuts as used in \citet{Trick18:Gaia}. Specifically, we enforce parallax uncertainties of $\delta \varpi/\varpi < 0.05$ and $\delta v_{\tot} < 8\,\kms$, where $v_{\tot}$ is the total velocity of a star.\footnote{See \citet{Trick18:Gaia} for a convenient $v_{\tot}$ formula.} These cuts ensure that our action estimates are accurate. We make a further cut, enforcing that the flag \verb|astrometric_excess_noise_sig| be less than $2$. This excludes sources with potentially extended profiles (e.g. binaries, galaxies), for which parallax and proper motion measurements are prone to systematics \citep{GAIAastrometry18}. After making these cuts, we retain 78 solar twins and 4376 stars from the APOKASC2 catalog.

We use the publicly available package \texttt{gala} to compute actions \citep{gala,Price-Whelan:2018}. In \texttt{gala}, the action integrals are evaluated numerically based upon an action estimator technique presented by \citet{Sanders14:actioncalc}. For the potential, we make use of the default \texttt{MWPotential}. This potential includes a Hernquist bulge and nucleus \citep{Hernquist90}, a Miyamoto-Nagai disk \citep{Miyamoto75}, and an NFW halo \citep{NFW97}, and is fit to empirically match some observations. This potential was based on the Milky Way potential available in \texttt{galpy} \citep{Bovy15:galpy}. We use the Dormand-Prince 8(5,3) integration scheme \citep{Dormand80:integrator} included in \texttt{gala}. We use a timestep of $1\,\Myr$ and integrate for $5\,\Gyr$, corresponding to $\sim 20$ orbits for a Sun-like star. To compute action errors, we performed 300 Monte Carlo realizations on the proper motion, radial velocity and parallax uncertainties, assuming errors are distributed in a multi-variate normal distribution with covariances as provided in the {\em Gaia DR2} data.

\section{Results}\label{sec:results}

\subsection{Sample Characterization}
We begin by describing the actions, velocities, metallicities and $\alpha$-enhancements of our two samples of stars. 

Fig.~\ref{fig:char} ({\em upper left panel}) shows $V$ as a function of $U$. The heatmap shows the distribution for the APOKASC2 sample, while the open circles show the solar twin sample. Neither of our samples are complete enough to reliably discriminate any sub-structure in $V$-$U$ space, as previous authors have shown \citep{GAIA18:kinematics,Trick18:Gaia}. We do see a slight correlation between $U$ and $V$; stars with higher $V$ tend to also have higher $U$.

\begin{figure*}[t!]
\centering
\includegraphics[width=\textwidth]{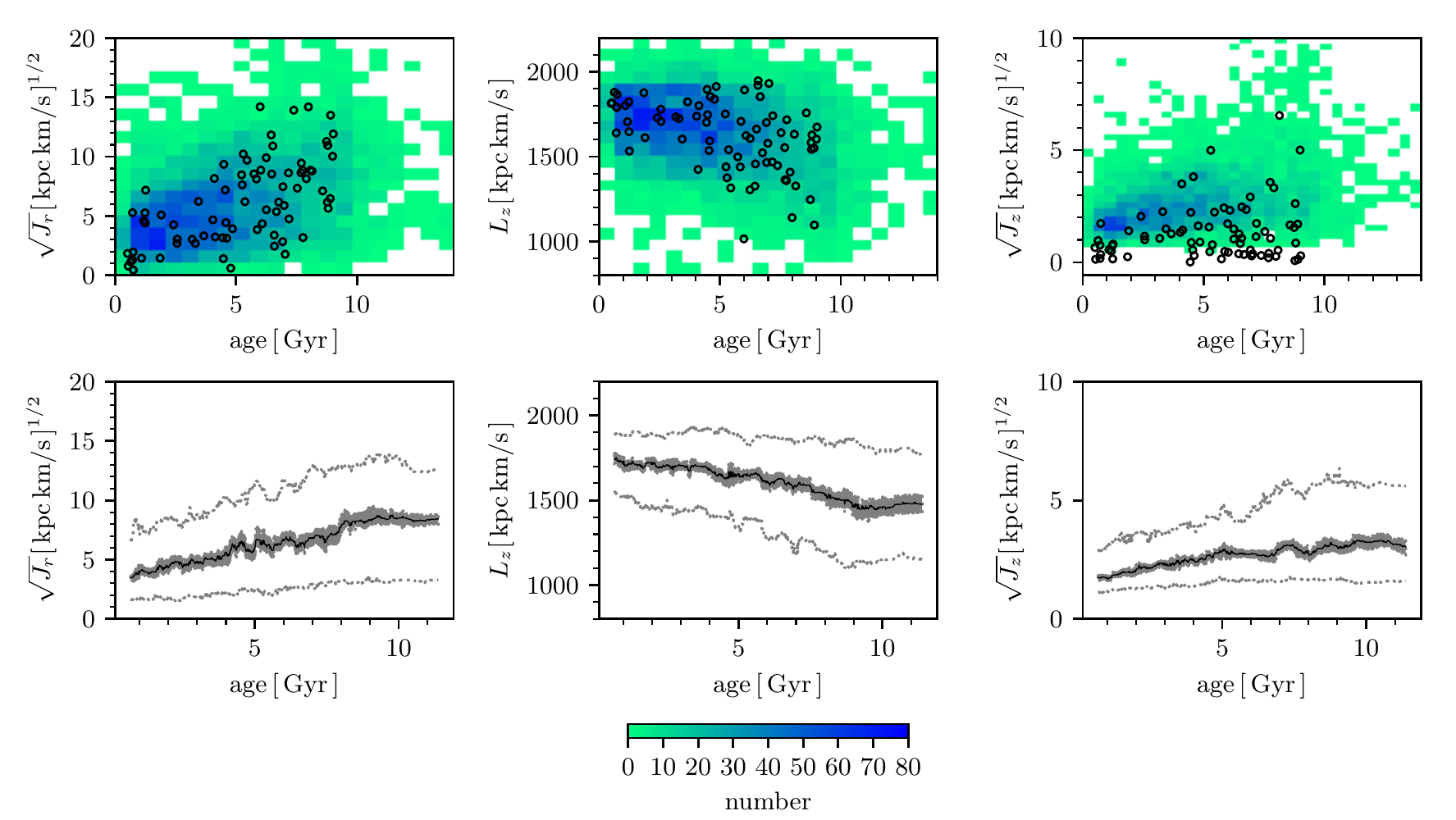}
\caption{The three actions versus age. The {\em upper panels} show the distribution of our two samples in action-age space. The heatmap corresponds to the APOKASC2 sample, while the foreground scatter plot shows the solar twin sample. The {\em lower panels} show the running median of each action vs. age in bins of 200 stars. The solid black line shows the running medians. 
To estimate the $2\text{-}\sigma$ error on the running medians, we bootstrap resample each bin of 200 stars 2000 times. This error is shown as the thickness of the running median lines. To depict the spread, we plot the 10\ts{th} and 90\ts{th} percentiles of each bin as dashed grey lines.
The distributions are fairly smoothly varying as a function of age. We compute the gradients to be $(7.51\pm0.52\text{,}\, {-29.0}\pm1.83\text{,}\,1.54\pm0.18)\, \actunit/\Gyr$ for $J_r$, $L_z$, and $J_z$. We caution that our computed gradients are influenced by the spatial bias in the APOKASC2 sample --- only $15\%$ of the APOKASC2 stars are at vertical heights $<200\,\pc$. Specifically, our computed gradient for $J_z$ is most strongly affected as stars with low $J_z$ will lie at low vertical heights, though we expect the results for all actions to be affected in detail.}
\label{fig:acts_vs_age}
\end{figure*}

Fig.~\ref{fig:char} ({\em upper right panel}) also shows $\sqrt{J_r}$ as a function of $L_z$.
The apparent cone-like nature of this plot is simply a selection bias. Stars with lower $L_z$ (and thus small orbital radius) and also low radial action $J_r$ than within the distribution of stars in this panel do not make excursions to the solar neighborhood, and are thus not included in our solar twin sample or our APOKASC2 sample (which is mostly within $300\,\pc$ of $R_{\sun}$).
The asymmetry in this plot is also well-understood and present for two reasons. First, there are simply more stars with lower $L_z$ (more stars near the Milky Way center). Second, nearby stars with low $L_z$ must be at apocenter, whereas nearby stars with high $L_z$ must be at pericenter. Since stars spend a larger fraction of their orbit at apocenter than at pericenter, this further explains the asymmetry in the {\em upper right panel} of Fig.~\ref{fig:char} (see also discussion in \citet{Trick18:Gaia}).

The {\em bottom right panel} of Fig.~\ref{fig:char} shows the APOKASC2 and solar twin samples in the $\alphafe-\feh$ plane. The solar twin sample is confined to a small space in $\feh$, reflecting the fact that they were chosen in part to be within $\sim 0.1\,\text{dex}$ of solar metallicity. The APOKASC2 sample shows the well-known metal-poor, $\alpha$-enriched branch, which is only marginally present in the relatively metal-rich solar twin sample \citep{Bensby03:alphabimodal,Bensby05:alphabimodal,Nidever14:alphabimodal,Hayden15:chemicalcartography}. 
This shows that the APOKASC2 sample includes both the low- and high-$\alpha$ sequence, which is referred to as the thin and thick disk in the solar neighborhood, respectively.

The spatial distribution of each sample is also shown in Fig.~\ref{fig:char} ({\em bottom left panel}). The solar twin sample is all within $100\,\pc$ of the Sun. The APOKASC2 sample is more broadly distributed extending to vertical heights of $\sim1\,\kpc$. Importantly, this sample is biased toward stars with current heights $z \gtrsim 200\,\pc$. Thus, we do not expect our results for the vertical actions to be representative for stars lying close to the galactic plane. The APOKASC2 sample has a radial extent spanning from $\sim7.5-8.1\,\kpc$. Because of this distribution, our results are representative of the correlations between actions and ages across a limited vertical extent ($0.3\pm 0.16\,\kpc$) and radial extent (of $7.8\pm0.11\,\kpc$), across the disk.  These correlations we find may change across the disk and we expect vastly different correlations extending to the Galactic halo, which has had a very different formation history to the disk \citep{Eggen62:halostarformation,Searle78:halostarformation}.
\begin{figure*}[p]
\centering
\includegraphics[width=\textwidth]{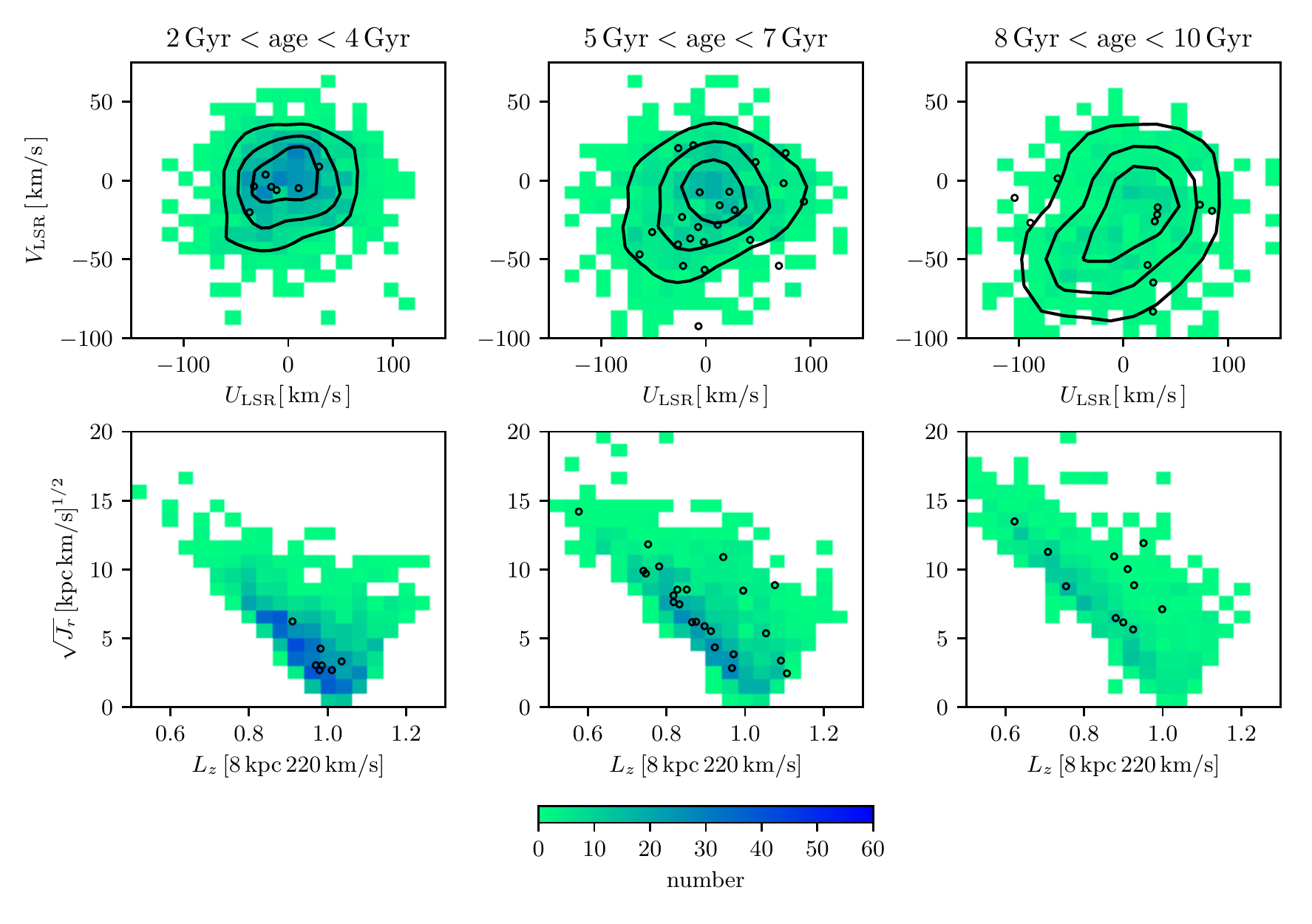}
\caption{Distribution of the solar twin sample (open circles) and the APOKASC2 sample (heatmap) for different age slices. The {\em upper panels} show the distribution in the $V$-$U$ plane, while the {\em lower panels} show the distribution in the $J_r$-$L_z$ plane. The contours in the {\em upper panels} enclose $25\%$, $50\%$, and $75\%$ of the APOKASC2 sample. The {\em left panels} show stars with ages between $2\,\Gyr$ and $4\,\Gyr$, the middle panels between $5\,\Gyr$ and $7\,\Gyr$, and the {\em right panels} between $8\,\Gyr$ and $10\,\Gyr$. This figure shows that while young stars are somewhat localized in kinematic and action space, old stars are uniformly distributed. Because of this, ages cannot be inferred from a star's position in action space. An animated version of this figure is available \href{https://gusbeane.github.io/actions-weak-age}{online}. Animation (6 seconds): Instead of six panels, we show two panels where the {\em upper panel} depicts the V-U plane and the {\em lower panel} depicts the $J_r$-$L_z$ plane. Each frame of the video shows a different age slice of $2\,\Gyr$. The age slice of each frame is shown by a bar on a new right-hand axis.}
\label{fig:JrLz_uv_time}
\end{figure*}

\begin{figure*}[p]
\centering
\includegraphics[width=\textwidth]{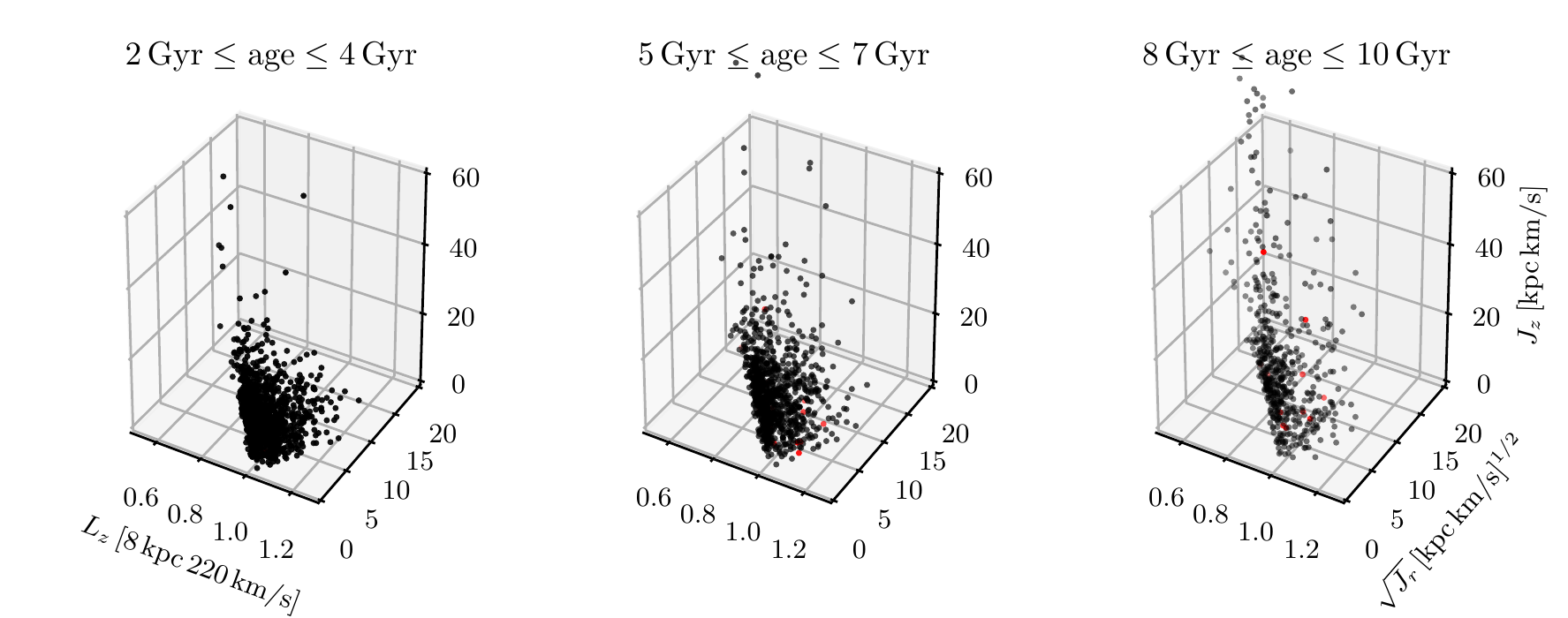}
\caption{A reproduction of the {\em lower panels} of Fig.~\ref{fig:JrLz_uv_time}, except now including the full 3D action space. The black scatter points correspond to the APOKASC2 sample, and the red points to the solar twin sample. The additional consideration of $J_z$ is not sufficient to discriminate between young and old stars in action space. As before, young stars are localized around circular orbits, whereas old stars are uniformly distributed in action space. An animated version of this figure, in which each subplot rotates, is available \href{https://gusbeane.github.io/actions-weak-age}{online}. Animation (14 seconds): Each subpanel of this plot rotates, to better show the 3D distribution of the stars in action space.}
\label{fig:JrLzJz_time}
\end{figure*}

\subsection{Action-Age Space}

We first examine the three actions versus age in Fig.~\ref{fig:acts_vs_age}. The heatmap in the {\em top panels} shows the number of stars from the APOKASC2 sample, while the open circles show the individual stars of the solar twin sample.
The {\em bottom panels} show the running medians (black lines) of each of the three actions as a function of age. The gray shading around the median is the $2\text{-}\sigma$ error calculated by bootstrap resampling each bin 2000 times, and the dashed gray lines show the 10\ts{th} and 90\ts{th} percentiles of each bin of stars to reflect the spread. The running medians and the associated errror and spread are calculated across bins of 200 stars. We performed a $6\text{-}\sigma$ clip on the total sample for each action, excluding a total of 70 stars. To compute gradients throughout, we measure the mean action $J_i$ for stars falling in two age bins: $0\text{--}2\,\Gyr$ and $9\text{--}11\,\Gyr$. After testing for convergence, we performed 5000 Monte Carlo samples on the age and action errors of all stars to determine the error on the gradient. We assumed the errors have a normal distribution.

The {\em upper left panel} of Fig.~\ref{fig:acts_vs_age} shows $\sqrt{J_r}$ as a function of stellar age and the {\em bottom left panel} shows the running median, error, and spread for this distribution. We plot $\sqrt{J_r}$ as a way to better visualize the large range in $J_r$. The radial action is lowest for young stars, the youngest of which are concentrated at very low values of $J_r$, and is largest for the oldest stars. The rate of change of $J_r$ as a function of age is $7.51\pm0.52\,\actunit/\Gyr$. 
Over the $1\text{-}10\,\Gyr$ timescale, $J_r$ increases by a factor of 3.4. The dispersion around the median\footnote{As measured by one half the difference between the 90\ts{th} and 10\ts{th} percentiles.} is large, about $55-80\%$ of the median value itself, decreasing as a percentage with age. The relatively weak gradient in $J_r$ coupled with the high variance implies a large overlap in the action distributions of stars as a function of age. Pivoting around the mean stellar age of $5\,\Gyr$, the entire solar twin sample and $93\%$ of the APOKASC2 sample have a $\sqrt{J_r} < 10 \sqrt{\actunit}$ for ages $<$ 5 Gyr. For stars with ages $>$ 5 Gyr, $76\%$ of the stars have $\sqrt{J_r} < 10\,\sqrt{\actunit}$.

The {\em upper middle panel} of Fig.~\ref{fig:acts_vs_age} shows the angular momentum, $L_z$, as a function of age. The {\em bottom middle panel} of Fig.~\ref{fig:acts_vs_age} shows the corresponding running median, error, and spread. Conversely to the radial action, the youngest stars have the highest values of $L_z$ and the oldest stars the lowest values. The angular momentum fairly smoothly decreases by a factor of $0.86$ from $1-10\,\Gyr$. The gradient is about ${-29}\pm1.83\,\actunit/\Gyr$.
The spread is on the order of $\sim10-25\%$ of the median value of $L_z$ at a given age, increasing from $\sim10\%$ for young stars to $\sim25\%$ for old stars. Among stars younger than $5\,\Gyr$, all but one of the solar twins and $85\%$ of the APOKASC2 sample have $L_z > 1500\,\actunit$. For stars with ages $> 5\,\Gyr$, $60\%$ have $L_z > 1500\,\actunit$. The oldest stars have angular momenta as low as $\sim 1000\,\actunit$; however, stars at all ages are seen at the lowest values of $L_z$. 

The {\em upper right panel} of Fig.~\ref{fig:acts_vs_age} shows the vertical action, $J_z$, as a function of age. Similar to $J_r$, we plot $\sqrt{J_z}$ to better visualize the large range in $J_z$. For this panel, the distribution of the solar twins in action space is different from that of the APOKASC2 sample. This is a consequence of the different spatial extent in $z$ for these samples --- Fig.~\ref{fig:char} ({\em bottom right panel}).
The solar twin sample resides within $100\,\pc$ of the Sun and stars with high vertical actions spend a smaller fraction of their orbit close to the galactic plane. Therefore, stars with high values of $J_z$ are less likely to reside within $100\,\pc$ of the Sun. The APOKASC2 sample, however, is located further from the plane, at a height of $0.35\pm0.16\,\kpc$. For this action it is clear that the spatial selection of our stars could bias our results, which we discuss further in \S~\ref{sec:discussion}.  Similarly to the radial action, the vertical action, $J_z$, increases with stellar age. The approximate rate of increase is $1.54\pm0.18\,\actunit/\Gyr$, which corresponds to an increase in the median $J_z$ value by a factor of 3.4 from 1 to 10 Gyrs. Pivoting around the mean age of $5\,\Gyr$, we see that $80\%$ of stars with ages $< 5\,\Gyr$ have $J_z < 10\,\actunit$, and  $57\%$ of stars older than $5\,\Gyr$ have the same $J_z < 10\,\actunit$. The dispersion around $J_z$ is on the order of $50-70\%$ of the median value itself, increasing with age with an increase at $\sim 6\,\Gyr$. Of potential curiosity is a bump in the $J_z-\text{age}$ relation at $\sim 7.5\,\Gyr$. However, the evidence for the bump in this sample is weak.

\begin{figure*}
\centering
\includegraphics[width=\textwidth]{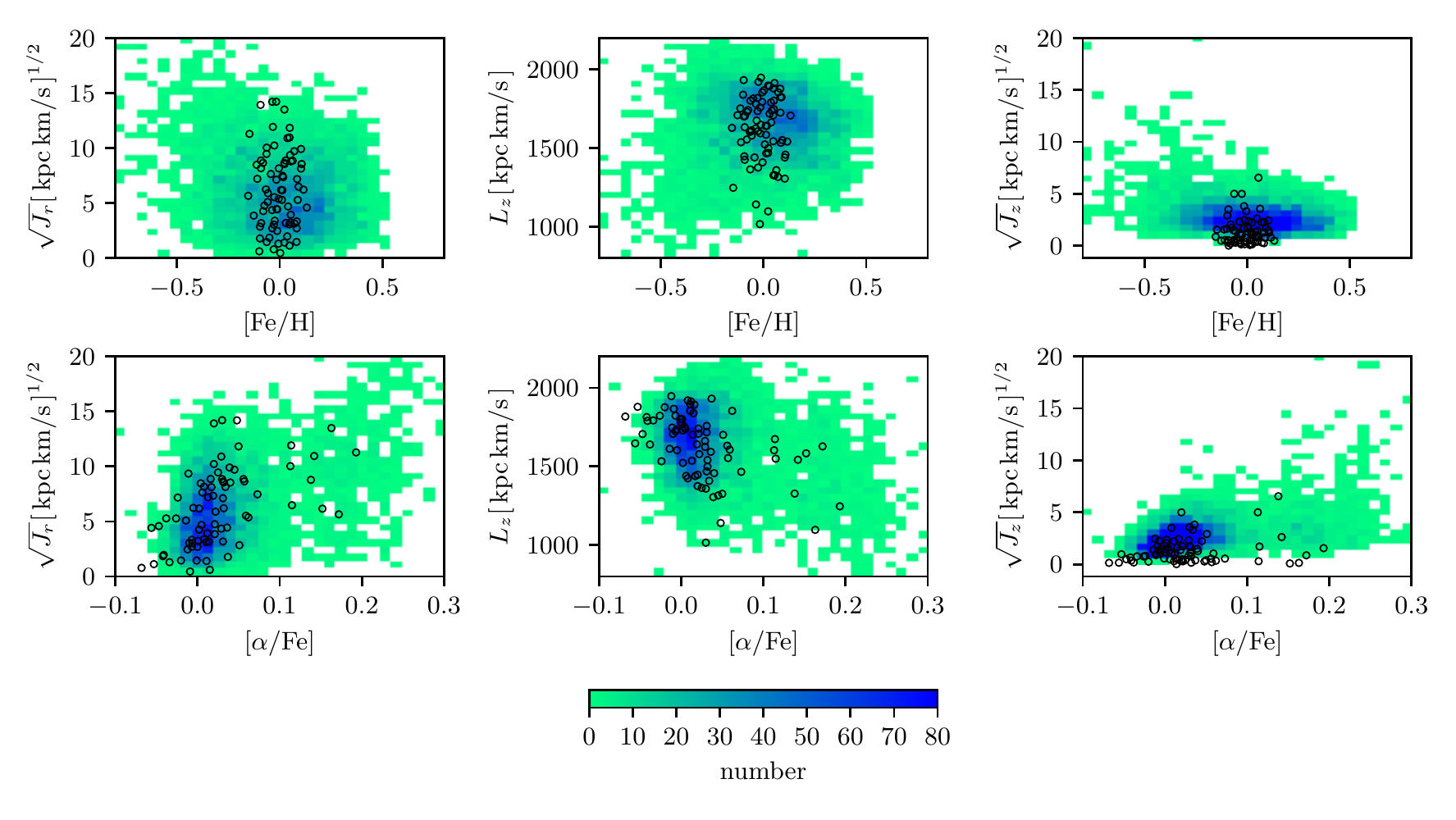}
\caption{The three actions vs. metallicity. The {\em upper panels} show the three actions as a function of metal enrichment $\feh$. The {\em lower panels} show the three actions as a function of $\alpha$-enrichment $\alphafe$. For the solar twin sample, the $\alpha$-enrichment indicated on the x-axes is actually a measured $\mgfe$ value. The {\em lower panels} show that while $\alpha$-enriched stars tend to have more non-circular and non-planar orbits, there is enough scatter that the $\alpha$-enrichment cannot be inferred from stellar actions in our sample.}
\label{fig:action_metal}
\end{figure*}

\subsection{Ages in Action and Velocity Space}\label{ssec:ages_in_act}
We now examine the age distribution in 2D planes of actions and in the velocity plane. Our goal here is to see if stars of a certain age are localized anywhere in action space, as opposed to just finding action-age correlations as in the previous sub-section.

We first explore the $V$-$U$ and $J_r$-$L_z$ planes in Fig.~\ref{fig:JrLz_uv_time}, for which an animated version is available \href{https://gusbeane.github.io/actions-weak-age}{online}. The upper three panels show the age distribution in $V$-$U$ space for different slices in age. The title of each panel indicates which age slice is being shown in that panel. The {\em left panels} show the number densities for stars with ages between $2\,\Gyr$ and $4\,\Gyr$, the {\em middle panels} for ages between $5\,\Gyr$ and $7\,\Gyr$, and the {\em right panels} for ages between $8\,\Gyr$ and $10\,\Gyr$. The upper three panels show that while young stars do show some localization in $V$-$U$ space, intermediate-aged and older stars are more uniformly distributed.

Similarly, we plot the number densities in $J_r$-$L_z$ space for different slices in age in the bottom panels of Fig.~\ref{fig:JrLz_uv_time}. The {\em left panel} shows how young stars are localized around low radial actions, with the entire solar twin sample and $85\%$ of the APOKASC2 sample satisfying $\sqrt{J_r} < 8\,\sqrt{\actunit}$. In the {\em middle panel}, we see that intermediate aged stars have a much wider distribution in $J_r$-$L_z$ space, and similarly for the old stars in the {\em right panel}. Perhaps the lowest radial actions are slightly disfavored for the oldest stars, though our samples are not large enough to be certain of this.

One might wonder whether the degeneracy between old and young stars in the $J_r$-$L_z$ plane can be broken by considering also the vertical action $J_z$. This is not the case. Fig.~\ref{fig:JrLzJz_time} reproduces the {\em lower panels} of 
Fig.~\ref{fig:JrLz_uv_time}, except now in full 3D action space. The vertical axis corresponds to the vertical action $J_z$. These panels are consistent with our previous interpretations. Young stars are localized around circular orbits in action space, but old stars are more uniformly distributed. This again indicates that old stars exhibit both circular- and elliptical-like orbits. An animated version of Fig.~\ref{fig:JrLzJz_time}, in which each subplot rotates, is available \href{https://gusbeane.github.io/actions-weak-age}{online}.

\subsection{Action-Abundance Space}\label{ssec:actmetal}
We next examine action-metallicity space in Fig.~\ref{fig:action_metal}.
The upper three panels show the three actions vs. $[\text{Fe}/\text{H}]$. As expected, the scatter plot of the solar twins are clustered around solar metallicities. One can see that there are no strong correlations between any of the three actions and $\feh$. The variance of the actions is however higher for the more metal-poor stars ($\feh \leq -0.3 \text{dex}$) in our sample. 

The bottom three panels of Fig.~\ref{fig:action_metal} show how the three actions correlate weakly with $\alpha$-enrichment. The $\alpha$-enriched stars do exhibit higher radial actions, but there are still an abundant number of $\alpha$-poor stars with similarly high radial actions. Similarly, while $\alpha$-enriched stars tend to have lower angular momenta $L_z$, there is still a number of $\alpha$-poor stars with similarly low $L_z$. What this means is that the $\alpha$-enrichment of a star cannot be accurately determined from either $J_r$ or $L_z$. Although $\alpha$-enriched stars have, on average, higher vertical actions, $J_z$, compared to more $\alpha$-poor stars, the dispersion in $J_r$ at a given $\alphafe$ is large. Therefore, many $\alpha$-enriched stars have comparable values of $J_z$ to $\alpha$-poor stars.
However, stars with very high vertical actions tend to have higher $\alpha$-abundances, (although these may be halo and not disk stars). We highlight and caution that this result is influenced strongly by the narrow coverage across vertical height, $z$, of our stars. This narrow range in $z$ examines only a narrow distribution in $J_z$. 

\begin{figure}
\centering
\includegraphics[width=\columnwidth]{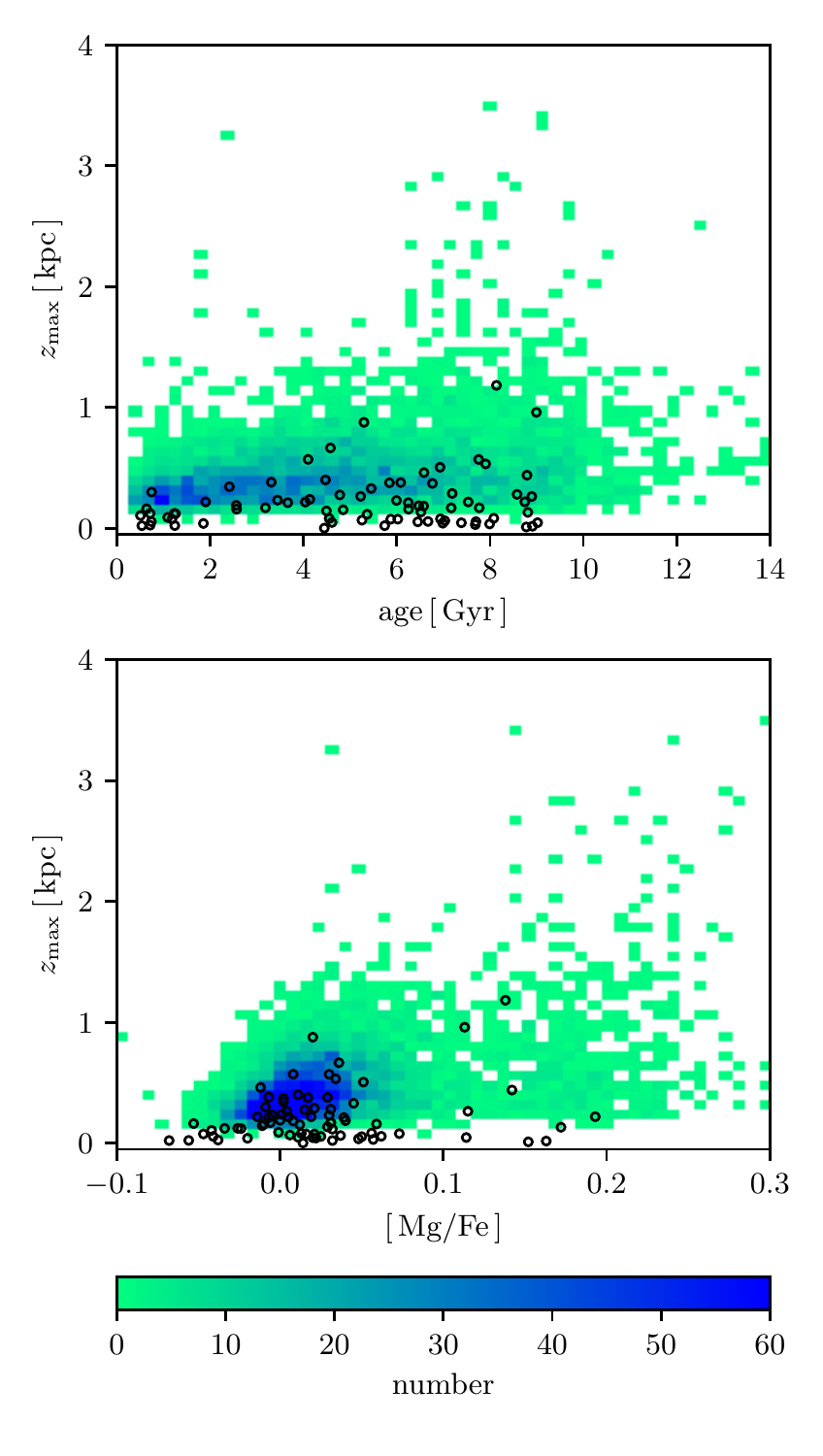}
\caption{The maximum vertical excursion, $z_{\text{max}}$, as a function of age ({\em upper panel}) and $\alpha$-enrichment ({\em lower panel}) for our two samples, with the heatmap representing APOKASC2 and the foreground scatter plot the solar twins. Note that while there is a general trend of older stars having greater vertical excursions, there is significant scatter about this relation. Because of the survey selection, the APOKASC2 sample only contains 27 stars with current heights $z<100\,\pc$. Thus, there are practically no APOKASC2 stars with $z_{\text{max}}<100\,\pc$, while there are many solar twins with such $z_{\text{max}}$.}
\label{fig:zmax_age}
\end{figure}

Fig.~\ref{fig:zmax_age} explores the $J_z$-metallicity relationship further by plotting the maximum vertical excursion of the APOKASC2 and solar twin stars as a function of age in the {\em upper panel}, and as a function of $\alpha$-enrichment in the {\em lower panel}. The {\em upper panel} confirms what we saw in the {\em right-upper panel} of Fig.~\ref{fig:acts_vs_age} --- young stars in our sample tend to be more localized while old stars are more broadly distributed, with significant scatter in the relation. The {\em lower panel} confirms what we saw in the {\em lower-right panel} of Fig.~\ref{fig:action_metal} --- that $\alpha$-enriched stars are not distinct in their actions compared to $\alpha$-poor stars. We note again that our bias in testing only a narrow height in $z$ above the galactic plane prevents us from making quantitative statements. For this it is essential to take into account the APOKASC2 selection function.

\section{Discussion}\label{sec:discussion}

In \S~\ref{ssec:ages_in_act} (Fig.~\ref{fig:acts_vs_age}) we found that older stars have higher orbital actions in $J_r$ and $J_z$ (and lower angular momentum, $L_z$). The gradient we find in $L_z$ is also consistent with inside-out formation of the Milky Way; the old stars reside in the inner Galaxy and the young stars in the outer. Of particular note is the large breadth of actions for both young and old stars. Already by $1\,\Gyr$, several stars in the APOKASC2 sample have high $J_r$, and several stars older than $10\,\Gyr$ have quite low radial actions. In Fig.~\ref{fig:age_hist}, we can see that stars younger than $1\,\Gyr$ have precise ages. However, it is still possible that unknown systematics have mischaracterized these stars' ages.

The fact that these stars exist is selection function independent, and indicates that radial mixing can occur on fairly short time scales (since some stars $<1\,\Gyr$ have high $J_r$). However, this is not guaranteed to act on every star (since some stars $>10\,\Gyr$ have low $J_r$). This suggests that radial mixing is caused by a few large events (e.g. spiral arm and bar resonances) as opposed to many small events (e.g. molecular cloud interactions). If radial mixing occurred over many small events, one would expect that: (i) there would not be a large spread in $J_r$ at a given age, and (ii) there would be very few old stars with low $J_r$. We see neither of these characteristics in Fig.~\ref{fig:acts_vs_age}. However, we caution that because our APOKASC2 sample preferentially contains stars with high vertical actions, the orbits of APOKASC2 stars will intersect the galactic plane less than a typical orbit and therefore will have fewer gas cloud interactions. Thus, quantification of this statement is beyond the scope of this work due to the complicated nature of the APOKASC2 selection function.

The significant spread in the relationships between the three actions and ages suggest that there is no clear way to distinguish young and old stars in action space. Nonetheless, it is tempting to use our observed localization of young stars in action space as an age proxy. Therefore, we investigate how well one can recover stellar ages from actions.

\begin{figure*}[t]
\centering
\includegraphics[width=\textwidth]{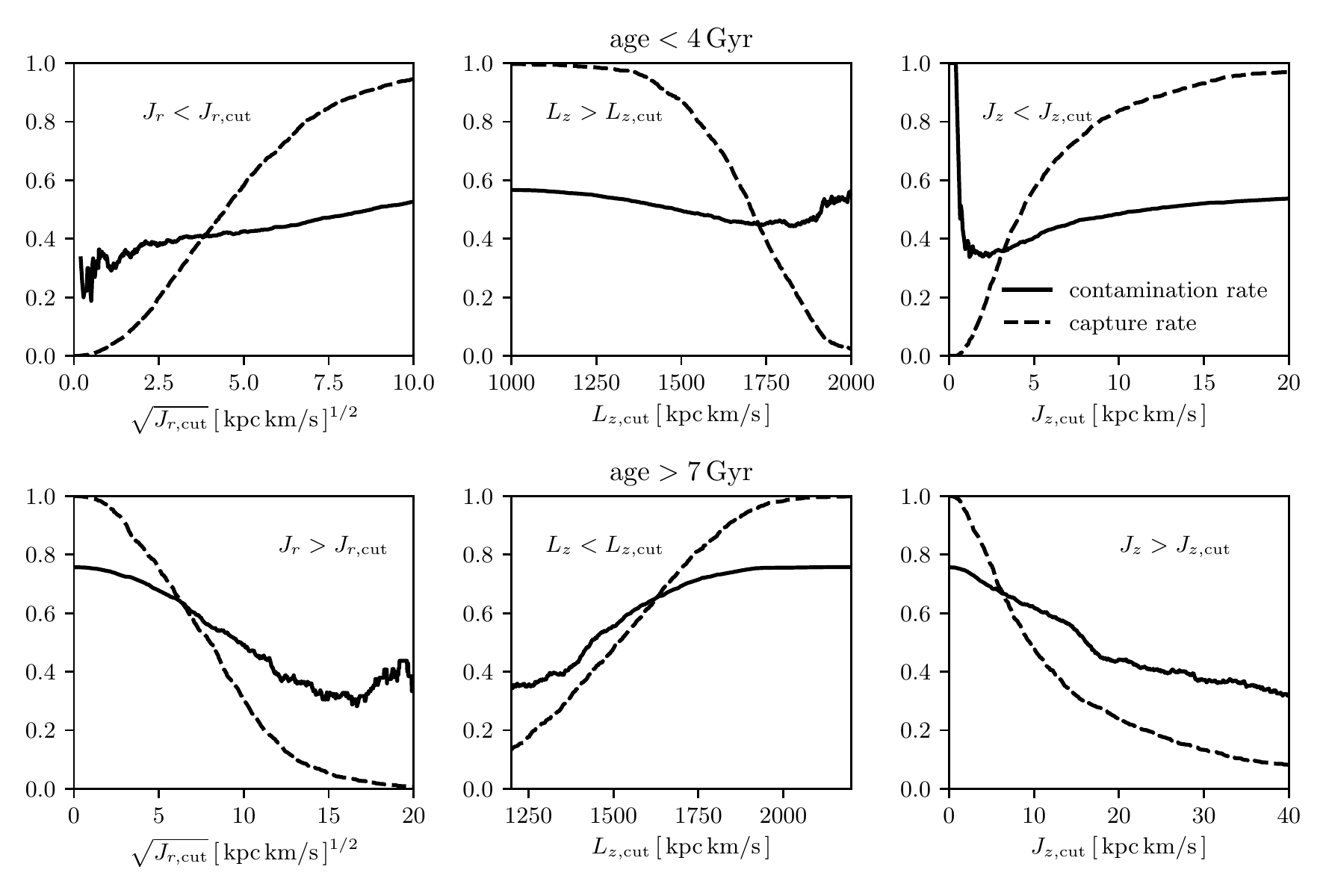}
\caption{The contamination and capture rate while attempting to separate young stars ($<$ 4 Gyr, {\em upper panels}) and old stars ($>$ 7 Gyr, {\em lower panels}). In the {\em upper panels}, we enforce action cuts which should correspond to younger stars: $J_r < J_{r,\text{cut}}$, $L_z > L_{z,\text{cut}}$, $J_z < J_{z,\text{cut}}$. We make the opposite cuts to separate older stars ({\em lower panels}). The contamination rate is the fraction of stars satisfying an action cut but are either older than $4\,\Gyr$ ({\em upper panels}) or younger than $7\,\Gyr$ ({\em lower panels}). Similarly, the capture rate is the fraction of stars younger than $4\,\Gyr$ ({\em upper panels}) or older than $7\,\Gyr$ ({\em lower panels}) that satisfy the given action cut. See Eq.~\ref{eq:captcontam} and accompanying discussion for details. At capture rates of $50\%$, we find contamination rates in young stars ({\em upper panels}) of $42\%$, $45\%$, and $39\%$ for cuts on $J_r$, $L_z$, and $J_z$, respectively. For old stars, we find at a capture rate of $50\%$ contamination rates of $56\%$, $57\%$, and $62\%$ for $J_r$, $L_z$, and $J_z$, respectively. This shows that no matter what action cut is made, we find high contamination rates.}
\label{fig:actcuts}
\end{figure*}

We only attempt to separate stars into two age bins: young stars ($<4\,\Gyr$) and old stars ($>7\,\Gyr$). These bins are arbitrary, but our results are not sensitive to the exact age cut used. We begin by filtering all stars satisfying a certain action cut --- for example, all stars with $J_r < J_{r,\text{cut}}$. We then quantify how well this age cut does by reporting, e.g., the fraction of stars younger than $4\,\Gyr$ satisfying that action cut, which we refer to as the ``capture rate''. Similarly, we also report the ``contamination rate'', or the fraction of stars that satisfy the action cut but are older than $4\,\Gyr$. For the case of separating young stars based on a $J_r$ cut, these are calculated as:
\beq\label{eq:captcontam}
\begin{split}
	\text{capture rate} &= \frac{N(J_r < J_{r,\text{cut}} \text{ and } \text{age}<4\,\Gyr)}{N(\text{age}<4\,\Gyr)} \\
    \text{contam. rate} &= \frac{N(J_r < J_{r,\text{cut}} \text{ and } \text{age}>4\,\Gyr)}{N(J_r < J_{r,\text{cut}})}\text{.}
\end{split}
\eeq

These are simply conditional probabilities. The capture rate quantifies how many of the total young stars are captured in an action cut sample while the contamination rate quantifies what fraction of the action cut sample are not young stars.
We do the same for each action, and also attempt to separate stars older than $7\,\Gyr$. For separating young stars, we enforce $J_r < J_{r,\text{cut}}$, $L_z > L_{z,\text{cut}}$, and $J_z < J_{z,\text{cut}}$, as suggested by Fig.~\ref{fig:acts_vs_age}. We perform the opposite cuts for separating old stars: $J_r > J_{r,\text{cut}}$, etc.

First, we consider separating out young stars ($<4\,\Gyr$) in Fig.~\ref{fig:actcuts} ({\em upper panels}). Making the cut $\sqrt{J_r} < 4.4 \,\sqrt{\actunit}$ gives a capture rate of $50\%$, meaning that $50\%$ of stars younger than $4\,\Gyr$ satisfy that $J_r$ constraint. However, at the same cut we find a contamination rate of $42\%$, meaning that $42\%$ of stars satisfying the cut in $J_r$ are older than $4\,\Gyr$. Similarly, we find a capture rate of $50\%$ for $L_z < 1706\,\actunit$, but a contamination rate of $45\%$. We find a capture rate of $50\%$ for $J_z < 4.4\,\actunit$, but a contamination rate of $39\%$.

The prospects for separating out old stars ($>7\,\Gyr$) are also shown in Fig.~\ref{fig:actcuts} ({\em lower panels}). Making the cut $\sqrt{J_r} > 8 \,\sqrt{\actunit}$, $L_z < 1514\,\actunit$, or $J_z >  9.5\,\actunit$ all give capture rates of $50\%$, but with contamination rates of $56\%$, $57\%$, and $62\%$, respectively. As can be seen in Fig.~\ref{fig:actcuts}, these high contamination rates are insensitive to the exact action cuts used, either for old or young stars.

\begin{figure}
\centering
\includegraphics[width=2.7in]{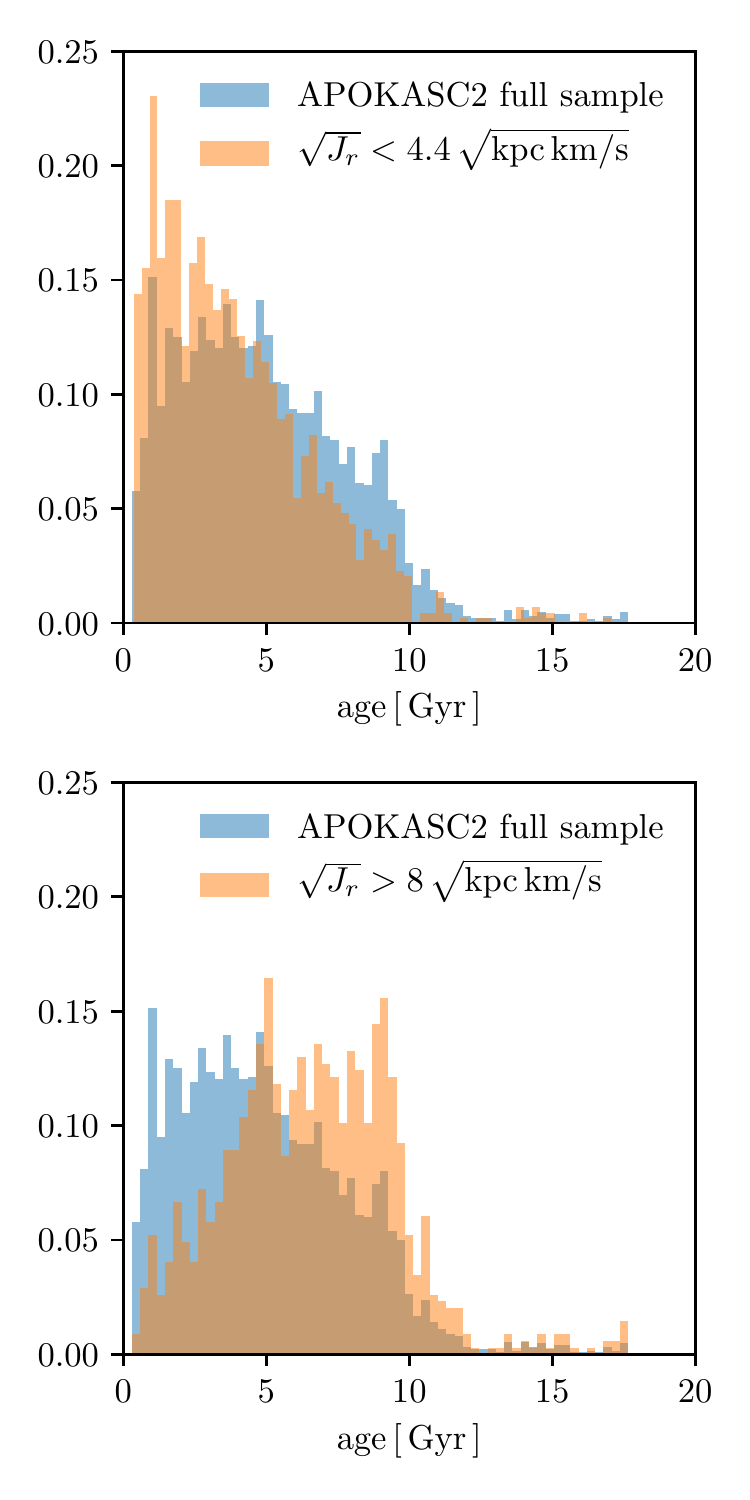}
\caption{Normalized histograms of the age distribution of the APOKASC2 sample we considered here (blue) as well as the age distribution of the subsample of stars satisfying a certain action cut (orange). We show the cut $\sqrt{J_r} < 4.4\,\sqrt{\actunit}$ ({\em upper panel}), which attempts to separate young stars, and $\sqrt{J_r} > 8\,\sqrt{\actunit}$ ({\em lower panel}), which attempts to separate old stars. The age distributions after making these action cuts are certainly more young ({\em upper}) or old ({\em lower}), but there is still significant spread in the age distribution after making each action cut. Each age distribution is similar to the APOKASC2 sample's distribution, suggesting that such an action cut might perform worse on a sample of stars with a more even age distribution.}
\label{fig:actcut_hist}
\end{figure}

We explore the age distribution of stars satisfying the $J_r$ cut in Fig.~\ref{fig:actcut_hist}, showing two overlaid normalized histograms for young stars ({\em upper panel}) and old stars ({\em lower panel}). The blue histogram shows the age distribution for the entire APOKASC2 subsample we consider in this work (reproducing the {\em lower panel} of Fig.~\ref{fig:age_hist}), while the orange histogram shows the age distribution for all stars satisfying $\sqrt{J_r} < 4.4\,\sqrt{\actunit}$ ({\em upper panel}) or $\sqrt{J_r} > 8\,\sqrt{\actunit}$ ({\em lower panel}). While in both cases the mean of the age distribution does shift, the effect is only marginal. Furthermore, the spread in the age distribution after making either action cut is too large for the action cut to be used as a reliable way to bin stars by age.

We showed in Fig.~\ref{fig:action_metal} that actions have weak correlations with $\alphafe$ (and even weaker trends for $\feh$). As with our action-age analysis however, these results are  influenced by the spatial extent of the APOKASC2 sample, warranting further study.  We found that $\alpha$-enrichment is however not a strong predictor of any of the three actions. Thus, action information cannot be used to infer individual stellar $\feh$ and $\alphafe$. 

Our results for vertical excursions in Figs.~\ref{fig:acts_vs_age}-\ref{fig:zmax_age} should not be affected by the expected flaring of the Milky Way, in which stars at larger galactic radii have larger vertical displacements. However, the solar twin sample is all within $100\,\pc$ of $R_{\sun}$, and $85\%$ of the APOKASC2 sample is within $300\,\pc$ of $R_{\sun}$ --- over such distances, flaring is not expected to be important \citep{Minchev12:quantradial,Ma17:flaring}.

As discussed in \S~\ref{sec:methods}, significant uncertainties in stellar ages can cause tight correlations to appear to be wider. However, excluding stars with age uncertainties more than $20\%$ does not qualitatively change our results. Furthermore, the scatter in Fig.~\ref{fig:acts_vs_age} is large enough to be robust to such an effect given the age uncertainties of our sample (see Fig.~\ref{fig:ageunc}).

We repeat that our results are sensitive to the limited spatial extent of our stars. The APOKASC2 sample is heavily biased against stars near the galactic plane --- only $15\%$ of the APOKASC2 stars have $z < 200\,\pc$. As a result, our sample does not reflect the underlying vertical distribution, age distribution, etc. This should propagate most strongly into our results for $J_z$, and less so for $J_r$ and $L_z$. We expect this to have a quantitative impact on the age gradient we found for $J_z$, though the observed variance in the $J_z$-age relation (Fig.~\ref{fig:acts_vs_age}) is expected to be qualitatively robust.

\section{Conclusions}\label{sec:conclusions}
Understanding the structure and evolution of the Milky Way disk requires an empirical characterization of how orbital properties of stars correlate with age \citep[see also][]{SA2018}.
As stars age, it is believed that their orbits become more non-circular and non-planar. These characteristics of an orbit are quantified in the actions $J_r$ and $J_z$, respectively. We considered these two actions, along with a star's orbital angular momentum, $L_z$, for two stellar samples with well-determined ages: a sample of $78$ solar twin stars and a $4376$ star subsample of APOKASC2. In summary, our conclusions are as follows: 

\begin{itemize}
\item We found weak correlations between actions and ages. We determined gradients of $(7.51\pm0.52\text{,}\, {-29.0}\pm1.83\text{,}\,1.54\pm0.18)\,\actunit/\Gyr$ for $J_r$, $L_z$, and $J_z$, respectively for our sample.
\item Old stars tend to be highly dispersed in action space, while young stars are more localized in action space, tending to have smaller $J_r$ and $J_z$ and higher $L_z$. We did not find that old stars {\em exclusively} have high $J_r$, $J_z$ or low $L_z$. This suggests that large radial mixing events occurring on short time scales (e.g. spiral arm and bar resonances) are more important than weaker radial mixing events occurring over longer time scales (e.g. molecular cloud interactions). We caution that this conclusion with respect to the relative importance of the two events is sensitive to the APOKASC2 selection function. Therefore further work is required to quantify the relative contribution and rate of different perturbation mechanisms.
\item We demonstrated that despite the presence of action-age gradients, there is significant spread in the distribution at a given age. There is significant overlap in the distribution of actions across the entire age range we considered, and as a result we concluded that inferring individual stellar ages from their actions is not possible.
We showed that even performing selecting either the youngest or oldest stars, respectively,  in action space,  results in contamination rates of $\gtrsim40\%$.
\item Despite the fact that actions cannot be used to infer stellar age alone, the action-age gradients we find indicates that dynamical actions are informative parameters for age inference \citep[e.g.][]{Sanders18:bayesages}.
\item We were unable to use actions to discriminate between high-$\alpha$ and low-$\alpha$ stars.
\end{itemize}

The entire solar twin sample was within $100\,\pc$ of $R_{\sun}$, and $85\%$ of the APOKASC2 sample was within $300\,\pc$ of $R_{\sun}$. As a result, our conclusions are only valid for the solar neighborhood. Further, the APOKASC2 sample was heavily biased against stars close to the galactic plane, with only $15\%$ of the sample within $200\,\pc$ of the galactic plane. Because of the complicated nature of the APOKASC2 selection function, here we restrict our work to be a general description of the overall correlations and description of the age-action correlations. Further work is necessary to understand what our results imply about secular evolution in the disk. We defer an analysis of the correlations between age and dynamics for the underlying population, that takes into account the selection function, for future work. Larger samples covering $z$ closer to the galactic plane will also help understand the action properties of the youngest stars in the Milky Way disk.

\section*{Supplementary Material}
All code and data used in this work is available at: \url{https://github.com/gusbeane/dyndat}. Please contact the author if you plan to use the data in a published work.
Animated versions of Figs.~\ref{fig:JrLz_uv_time}\,\&\,\ref{fig:JrLzJz_time} are available at: \url{https://gusbeane.github.io/actions-weak-age}.

\section*{Software}
This work made use of the following software: \texttt{astropy} \citep{astropy:2013, astropy:2018}, \texttt{gala} \citep{gala,Price-Whelan:2018}, \texttt{makecite} \citep{makecite:2018}, \texttt{matplotlib} \citep{Hunter:2007}, \texttt{numpy} \citep{numpy:2011}, and \texttt{scipy} \citep{scipy:2001}.

\section*{Acknowledgements}
We would like to thank David Spergel, Lorenzo Spina, Lauren Anderson, Kathryn Johnston, Stephen Feeney and David W. Hogg for helpful discussions. AB would like to thank Todd Phillips for helpful discussions. AB was supported in part by the Roy \& Diana Vagelos Program in the Molecular Life Sciences and the Roy \& Diana Vagelos Challenge Award.

This work has made use of data from the European Space Agency (ESA) mission
{\it Gaia} (\url{https://www.cosmos.esa.int/gaia}), processed by the {\it Gaia}
Data Processing and Analysis Consortium (DPAC,
\url{https://www.cosmos.esa.int/web/gaia/dpac/consortium}). Funding for the DPAC
has been provided by national institutions, in particular the institutions
participating in the {\it Gaia} Multilateral Agreement.

\bibliography{references}

\end{document}